\documentclass[aps,twocolumn,amsmath,amsfonts,amssymb,groupedaddress]{revtex4}

\usepackage{amsmath}
\usepackage{amsfonts}
\usepackage{amssymb}
\usepackage{stmaryrd}
\usepackage{graphics,epsfig}
\usepackage{bm}
\begin{document}

\title{Quantum Fine-Grained Entropy}

\author{Dong-Sheng Wang}
\email{wdscultan@gmail.com}
\date{4 May, 2012}

\begin{abstract}
Regarding the strange properties of quantum entropy and entanglement, e.g., the negative quantum conditional entropy, we revisited the foundations of quantum entropy, namely, von Neumann entropy, and raised the new method of quantum fine-grained entropy. With the applications in entanglement theory, quantum information processing, and quantum thermodynamics, we demonstrated the capability of quantum fine-grained entropy to resolve some notable confusions and problems, including the measure of entanglement and quantumness, the additivity conjecture of entanglement of formation etc, and the definition of temperature for single quantum system.
\end{abstract}

\maketitle

\section{INTRODUCTION}
\label{sec:1INTRODUCTION}

Entropy is one fundamental concept in physics and has been diversified in different studies, e.g., information processing and thermodynamics \cite{wehrl,peres,horodecki,Nielsen}. In these years, the development of the entanglement theory and quantum information processing (QIP) has stressed the significance of quantum entropy, e.g., von Neumann entropy and the quantum relative entropy. However, weirdness of quantum characters has caused much attention as well as trouble in the study of quantum entropy and entanglement. Discordance exists between the classical Shannon entropy and the quantum entropy. Also, due to entanglement, the property of quantum entropy becomes pretty strange, e.g., the quantum conditional entropy can be negative. Various measures of entanglement could confuse the physical essence of entanglement, also the difference between entropy and entanglement.

The difficulties in the study of quantum entropy and entanglement not only simply originate from the applications, also from some fundamental issues. One difficulty comes from the concept of entropy itself, e.g., there exists confusion whether entropy is objective or subjective, i.e., whether entropy describes the intrinsic uncertainty or our uncertain knowledge of the deterministic state. Confusions also come from whether it should be extensive or nonextensive. Another difficulty is from quantum mechanics. For instance, for one unmeasured system, what is its state? i.e., does it has one state while it is not measured? Are there two kinds of uncertainty, classical and quantum? These difficulties result in some notable problems, e.g., the relation between entropy and entanglement, the additivity of some quantities in QIP, the definitions of entropy and temperature for single quantum system etc. We will focus on these problems in our study, particularly, we address the concepts of ``coarse-graining'' and related ``fine-graining'', which are seldom noticed in the study of quantum entropy and entanglement, and we introduce in the concept of ``quantum fine-grained entropy'' to resolve several related problems and explore some interesting features of quantum entropy.

There are mainly three parts of this work. In section \ref{sec:2FGQE}, we introduce in the method of quantum fine-grained (QFG) entropy. In subsection \ref{sec:21graining}, we firstly discuss the physical basis for coarse-graining and fine-graining. To clarify the physical essence of QFG entropy, we investigate the formalism of state matrix in subsection \ref{sec:22reason}, we also argue that the existence of QFG entropy is indicated in the study of QIP and quantum foundation. We present the primary formalism of QFG entropy in subsection \ref{sec:23von}, and show that von Neumann entropy is a kind of coarse-grained entropy. In section \ref{sec:3prop}, we study the fundamental requirements for the definition of entropy, based on which, we analyze the main properties of Shannon entropy, von Neumann entropy and the QFG entropy, also highlight their difference. After the study of the foundation of QFG entropy, in section \ref{sec:4app}, we explore some of the potential applications, mainly in three aspects: entanglement theory in subsection \ref{sec:41entan}, quantum information processing (QIP) in subsection \ref{sec:42QIP}, and quantum thermodynamics in subsection \ref{sec:43statis}. At the end in section \ref{sec:5CONCLUSION}, we conclude.

\section{Quantum fine-grained entropy}
\label{sec:2FGQE}

In this section, we aim to introduce in the concept of quantum fine-grained entropy. The physical foundation of QFG entropy is based on the methods of coarse-graining and fine-graining, which relates to renormalization, also on the theory of density matrix. We also verify that the commonly used von Neumann entropy is a kind of coarse-grained entropy.

\subsection{Coarse-graining and fine-graining} \label{sec:21graining}

The method of coarse-graining is often involved in the theory of statistics and many-body dynamics \cite{coarse}. In the many-body system, there can be dynamics with different time and space scales simultaneously. For instance, the reaction between atoms involves the dynamics of electrons, as well as the dynamics of nucleus, which can be ignored via the Born-Oppenheimer approximation. The existence of various time and space scales results in the complexity of the dynamics. Coarse-graining is the method used to reduce the complexity of the dynamics also the geometric structure of the system, and sometimes it has to be employed since the fine-grained dynamics is not attainable. Generally, there are kinds of coarse-graining in practice, here we can roughly clarify two types according to the scale property of the dynamics. type-I: coarse-graining within one scale, e.g., ignoring and tracing out some bodies of the many-body dynamics. type-II: coarse-graining between scales, e.g., ignoring the fast process i.e. the effect of this process is time-averaged out. Note that our simple classification of the types of coarse-graining does not affect the validity of our investigation on the QFG entropy.

There exist some well-known coarse-graining procedures in Classical Mechanics (CM). The classical dynamics can be described in the phase space, which can be divided into lattice with finite action of each region. In practice, there can be different strategies for coarse-graining. The basic one comes from the uncertainty principle $\Delta p \Delta q \geq\hbar/2$ due to Quantum Mechanics (QM), which provides the fundamental limit of the precision of measurement. Also, the information we know about the dynamics at each point of the phase space might be only probably true, thus we have to make average of action of each region. Another strategy can be due to some of the degree of freedom (d.o.f) of the dynamics is totally uncertain, then we have to ignore the d.o.f we do not know. Note that, the difference between ignoring and tracing out is trivial for classical dynamics since measurement does not play the fundamental roles, which is not the case for QM studied below.

In QM, the dynamics of density matrix acts in the Hilbert space \cite{footnote1}. One of the well-known yet not well-understood coarse-graining is that in many-body system, e.g., multi-spins, the local state of one body is deduced by tracing out the other parts of the global states. However, coarse-graining does not directly mean tracing out, instead, there exists a step of ``ignoring'' beforehand. For instance, for Bell state $|\psi\rangle=\frac{\sqrt{2}}{2}(|00\rangle+|11\rangle)_{AB}$, firstly by ignoring state of $B$ we get $|\psi'\rangle=\frac{\sqrt{2}}{2}(|0\rangle+|1\rangle)_A|B\rangle$, where $|B\rangle$ labels state of $B$; then tracing out system $B$, we finally get $|\psi_A\rangle=\frac{\sqrt{2}}{2}(|0\rangle+|1\rangle)$, which is a superposed state. This is obviously different with the standard approach, where we trace out system $B$ directly resulting the totally mixed state $\rho_A=\textrm{diag}(1/2,1/2)$. Note that state $\rho_A$ is the result of the optimal measurement of state $|\psi_A\rangle$, yet, coarse-graining does not actually involve any measurement. This is type-I coarse-graining. There exists another coarse-graining type in Hilbert space, which is type-II, that we take several non-orthogonal states as orthogonal basis. For instance, for state $\rho=\sum_i p_i |\phi_i \rangle\langle \phi_i|$, $\langle \phi_i|\phi_j\rangle\neq\delta_{ij}$, apply coarse-graining as $|\phi_i\rangle$$\rightarrow$$|i\rangle$ with $\langle i|j\rangle=\delta_{ij}$, then state $\rho=\textrm{diag}(p_1, p_2, \cdots)$. This kind of coarse-graining can be understood as a mean of labeling or encoding. This type of coarse-graining is involved in the definition of von Neumann entropy, as we will analyze later on. Besides, there can be other kinds of coarse-graining in Hilbert space \cite{wehrl}, e.g., the density matrix can be cut off into blocks and some blocks are replaced by identity matrix while maintaining the trace.

In this work, instead of further analyzing on the method of coarse-graining itself, we will mainly focus on the interplay between quantum coarse-graining and quantum entropy.

\subsection{Why do we need quantum fine-grained entropy in quantum mechanics?} \label{sec:22reason}

To clarify the concept of quantum entropy, we need to start with the theory of density matrix, in the following, we will call $\rho$ as ``state matrix'' following the book \cite{wisemanbook} since the phrase ``density'' is quite misleading. For single system, the complete description of the system is the state vector $|\psi\rangle$, written in matrix formula as state matrix $\rho=|\psi\rangle\langle\psi|$, where the coherence (non-diagonal elements) is complete. If the coherence of the state matrix degenerates, we call the state as ``mixed state'', where $\rho$ cannot be written as $|\psi\rangle\langle\psi|$, if the coherence disappears, we call it as ``classical state''. One well-known fact is that the physical system corresponding to one state matrix $\rho$ is not unique, which indicates that the physical reality of a system relies directly on the detailed decomposition of the state matrix, as the result, the quantum entropy we can manipulate should be different.

It is established that the classical world results from decoherence and entanglement \cite{zurek03}, here we show that in addition to which, the classical state can directly come from classical mixing. To start, we present the formalism of state matrix \cite{Landau} in a new way. For many-body system, each body $i$ lives in one Hilbert space $\mathfrak{H}_i$. There can be two basic types of the composite space, one is in direct-sum form $\mathfrak{H}_I=\bigoplus_i^n\mathfrak{H}_i$, the other is in direct-product form $\mathfrak{H}_{II}=\bigotimes_i^n\mathfrak{H}_i$. For $\mathfrak{H}_I$, there must be no (or weak) interaction among the bodies, i.e., they form an ensemble, each body obeys the Liouville equation $i\hbar \dot{\rho}_i=[\mathcal{H}_i, \rho_i]$, ($\lambda$), with Hamiltonian $\mathcal{H}_i$, the whole system obeys $i\hbar \dot{\rho}=[\mathcal{H}_I, \rho]$, ($\alpha$), where $\mathcal{H}_I=\bigoplus_i^n \mathcal{H}_i$, $\rho=\bigoplus_i^n \rho_i$. The equation ($\alpha$) is equivalent with the set of equations ($\lambda$) for each body. Now, make the sum of the set of equations ($\lambda$), note that there might be several bodies with the same $\rho_i$, thus we have the probabilities $p_i$ for $\rho_i$. After some algebra, we get $i\hbar \dot{\rho}=[\mathcal{H}'_I, \rho]$, ($\beta$), where $\rho=\sum_i^{m<n} p_i \rho_i$, $\mathcal{H}_I'=\frac{1}{n}\sum_i^n\mathcal{H}_i$ is the average Hamiltonian. In practice, for the ensemble of identical particles, they have the same Hamiltonian, then $\mathcal{H}_I'=\mathcal{H}_i$. The equations ($\alpha$) and ($\beta$) are equivalent, the two forms $\bigoplus_i^n \rho_i$ and $\sum_i^{m<n} p_i \rho_i$ of $\rho$ are also equivalent, and the mixed state just comes from the classical mixing since there is no interaction among them \cite{footnote2}. For $\mathfrak{H}_{II}$, where there exist interactions among the bodies, the global system can be decomposed as $|\Psi\rangle=\sum_i \alpha_i |e^i_1 e^i_2 \cdots e^i_n \rangle$, with $|e^i_j\rangle$ as the basis of the $j$-th body, which can be a multi-partite entangled state. In QM, we need to pay attention to the notion of ``single system'', since it can be single body, e.g., one electron, or can even be the whole universe, which is highly a many-body system, due to entanglement. For single body system, the only possible origin of classicality comes from the interaction with other system, e.g., environment $E$, otherwise it stays as pure state. The interaction with the general form $\mathcal{H}=\sum_i S_i\otimes E_i$, where operator $S_i$ ($E_i$) acts on the system $S$ ($E$), can entangle the $S$ and $E$ together. That is, the local coherence within $S$ and $E$ transfers to the global shared coherence (entanglement). When only measuring the system $S$, we will find that $S$ is in mixed state. However, before any measurement, the system $S$ is not necessarily in mixed state (also see Appendix \ref{app:example}). Note that the system can stay pure when this interaction is not enough to generate entanglement, e.g., if the state of $E$ is classical, such as magnetic field, the coherence will stay within the system. For multi-partite entangled state, by tracing out some parts, the state of the remains can be in mixed state depending on the types of entangling \cite{horodecki}. To sum up briefly of the analysis above, relying on the two different structures of many-body Hilbert space and state matrix, there are two kinds of origins of classicality: mixing and entanglement.

One of the consequence of the above study is that, for instance, for state $\rho=\textrm{diag}(1/4, 1/4, 1/4, 1/4)$, we are not aware of its physical origin, e.g., it could be the single body system entangling with an environment, or two-body system entangling with other bodies, or a mixture of the Bell states, for different physical reality, the corresponding quantum entropies should be different, and the method of fine-graining is necessary.

In the study of QIP, there are also some hints for the existence of the QFG entropy, here we briefly discuss some of them as follows. Firstly, information is often classified into two types: classical and quantum (speakable or unspeakable). We say a quantum particle carries quantum information, e.g., the spin of one electron, even the electron is in a pure state $|\psi\rangle=\alpha|\uparrow\rangle+\beta|\downarrow\rangle$, it means that the entropy of pure state should not be zero! Secondly, different pure states are different, physically. For instance, the no-cloning theorem \cite{clone} states that it is impossible to clone unknown states, e.g., two non-orthogonal states $|\psi_1\rangle$ and $|\psi_2\rangle$. To clone one state, we need to know all the information of the state; thus, no-cloning means the information of them are different. Further, we know the no-cloning theorem is equivalent with state discrimination which means that non-orthogonal states cannot be perfectly distinguished, which is further the analogue to the classical case to distinguish two probability distributions $\{p_i\}$ and $\{q_i\}$. Obviously, there are entropies of the two distributions, then it is reasonable that the entropies of the states $|\psi_1\rangle$ and $|\psi_2\rangle$ are not zero, either. Thirdly, from the spectrum theory of matrix, the usual quantum entropy only addresses the eigenvalues. However, the eigenvalues and eigenvectors both have physical meanings, the eigenvalues cannot represent the properties of the matrix totally, e.g., the Pauli matrix have the same eigenvalues $\lambda_{1,2}=\pm1$, yet, they are different operations.

Furthermore, from the perspective of quantum foundation,  the basic elements in QM are algebra and vector space, different from CM. {\em Statistics} is inherent in quantum dynamics, which is a way to describe the statistical properties of reality, independent of deterministic description, e.g., CM or Relativity. That is to say, the {\em completeness} of quantum description of reality is in the quantum probabilistic sense. As a result, the wave function, although complete in the quantum sense, contains inherent uncertainty thus nonzero entropy.

From the above arguments, we can draw some lessons, for clarity, listed as follows.

{\em Lesson 1.} The quantum entropy of the pure state is not zero.

{\em Lesson 2.} The quantum entropies of the same $\rho$ with different decompositions are different. Thus, the entropy $\rho$ contains, the information we can possibly extract from it, and the work needed to ``erase'' the information depends on its decomposition.

{\em Lesson 3.} For many-body system, coarse-graining carries out by ignoring some parts instead of directly tracing them out.

\subsection{QFG entropy $\&$ von Neumann entropy} \label{sec:23von}

Next, we present the basic formalism of the QFG entropy. The state matrix of a general quantum system $\rho$ can be written as
\begin{equation}\label{eq:statematrix}
    \rho=\sum_i^n p_i \rho_i,
\end{equation}
we call each $\rho_i$ as a ``sector'', $n$ is the number of sectors, $\sum_i^n p_i=1$. The dimension of the system we study in this work is finite, the generalization to infinite dimension case is direct. Without lose of generality, we do not specify the dimension of the system also the normalization relations we study below unless necessary.

Next, we define the QFG entropy for pure state and mixed state.

For pure state $\rho=|\psi\rangle\langle\psi|$, the system can be single-body or many-body.

For single-body system, under the orthonormal basis $\{|k\rangle\}$, define $|\psi\rangle=\sum_k \alpha_k |k\rangle$. The QFG entropy is
\begin{equation}\label{qfg1}
    S_{FG}(|\psi\rangle)\equiv-\sum_k |\alpha_k|^2 \log |\alpha_k|^2= H(|\alpha_k|^2),
\end{equation}
where $H(\cdot)$ is Shannon entropy. The QFG entropy for pure single-body system, highly depending on the decomposition, is generally nonzero, which is different from the common result. Also, the QFG entropy does not quantify the effects of the relative phase difference among the basis, since the phase relates to coherence directly rather than entropy, thus the QFG entropy manifests the difference between coherence (also entanglement) and entropy.

Note that there also exists the non-orthogonal decomposition of pure state as $|\psi\rangle=\sum_l \gamma_l |\phi_l\rangle$, with $\langle \phi_l|\phi_{l'}\rangle\neq \delta_{l'l}$ \cite{wang1}. The QFG entropy is $S_{FG}(|\psi\rangle)=-\sum_l |\gamma_l|^2 \log |\gamma_l|^2+\sum_l |\gamma_l|^2 S_{FG}(|\phi_l\rangle)$. By type-II coarse-graining, set $\{|\phi_l\rangle\}$ as orthonormal basis, the QFG entropy reduces to that in Eq. (\ref{qfg1}).

For two-body entangled system $|\Psi\rangle=\sum_r a_r |\psi^r_A\psi^r_B\rangle$, the states $\{|\psi^r_A\rangle\}$ ($\{|\psi^r_B\rangle\}$) form the basis of $A$ ($B$). By type-I coarse-graining, we get $|\psi_A\rangle=\sum_r a_r |\psi^r_A\rangle$, $|\psi_B\rangle=\sum_r a_r |\psi^r_B\rangle$. The QFG entropy of the entangled system is
\begin{equation}\label{qfg2}
    S_{FG}(|\Psi\rangle)=H(|a_r|^2)=-\sum_r |a_r|^2 \log |a_r|^2,
\end{equation}
and $S_{FG}(|\psi_A\rangle)=S_{FG}(|\psi_B\rangle)=S_{FG}(|\psi\rangle)$. Then the mutual information between the two parts is
\begin{eqnarray}\label{qfgmut1}\nonumber
    I_{FG}(|\Psi\rangle)&=&S_{FG}(|\psi_A\rangle)+S_{FG}(|\psi_B\rangle)-S_{FG}(|\psi\rangle) \\
    &=&H(|a_r|^2),
\end{eqnarray}
which is totally quantum, thus we call it as {\em quantum correlation}. As a result, there is no classical correlation in the pure entangled two-body system.

For instance, for the bipartite entangled state $|\psi\rangle=\frac{1}{2}(|00\rangle+|10\rangle+|01\rangle-|11\rangle)$, which can also be written as $|\psi\rangle=\frac{\sqrt{2}}{2}(|0+\rangle+|1-\rangle)=\frac{\sqrt{2}}{2}(|+0\rangle+|-1\rangle)$, with $|\pm\rangle=\frac{\sqrt{2}}{2}(|0\rangle\pm|1\rangle)$, the QFG entropy of $|\psi\rangle$ is 1 bits, of each party is also 1 bits, thus the QFG mutual information is 1 bits.

For mixed state ($n\geq2$ in Eq. (\ref{eq:statematrix})), the QFG entropy is generally defined as
\begin{equation}\label{qfgrho}
    S_{FG}(\rho)=H(p_i)+\sum_ip_iS_{FG}(\rho_i),
\end{equation}
and the mutual information is defined as
\begin{equation}\label{qfgmut2}
    I_{FG}(\rho)=H(p_i)+\sum_ip_iI_{FG}(\rho_i),
\end{equation}
where $\rho_i$ can be mixed or pure. We can treat the part $H(p_i)$ in mutual information as ``classical'', and the remaining part can both contain classical and quantum correlations depending on the formulas of the sectors $\rho_i$. Note that the definition of QFG entropy does not depend on the $\rm{supp}(\rho_i)$, since the sectors are unconditional of each other.

For two-body system, the sector $\rho_i$ can be product state $\rho_i^A\otimes\rho_i^B$, or entangled state, or noise etc. For instance, for the separable state \begin{equation}\label{eq:separable}
 \rho=\sum_i p_i \rho_i^A \otimes \rho_i^B,
\end{equation}
by type-I coarse-graining, $\rho_A=\sum_i p_i \rho_i^A$, $\rho_B=\sum_i p_i \rho_i^B$, then the QFG entropies are
\begin{eqnarray}\label{}\nonumber
S_{FG}(\rho)&=&H(p_i)+\sum_i p_i S_{FG}(\rho_i^A)+\sum_i p_i S_{FG}(\rho_i^B), \\ \nonumber
S_{FG}(\rho_A)&=&H(p_i)+\sum_i p_i S_{FG}(\rho_i^A), \\
S_{FG}(\rho_B)&=&H(p_i)+\sum_i p_i S_{FG}(\rho_i^B),
\end{eqnarray}
and the mutual information is
\begin{equation}\label{}
I_{FG}(\rho)=H(p_i)=S_{FG}(\rho_A)+S_{FG}(\rho_B)-S_{FG}(\rho),
\end{equation}
which shows that the mutual information (correlation) in separable state is totally classical.

It is clear that the formalism of QFG entropy is distinct with von Neumann entropy. Note that the basic formulas of relative entropy and conditional entropy remains since their definitions do not involve coarse-graining.

In addition, there is one particular case that when there exist degenerate eigenvalues of the state matrix, e.g., for single-body completely mixed state $\rho=\textrm{diag}(1/2, 1/2)$. Obviously, under the basis $\langle0|=(1, 0)$, $\langle1|=(0, 1)$, it can be decomposed as $\rho=\frac{1}{2}\left( \begin{array}{cc} 1 & 0 \\ 0 & 0 \\ \end{array}\right)+\frac{1}{2}\left( \begin{array}{cc} 0 & 0 \\ 0 & 1 \\ \end{array}\right)$ with $S_{FG}=1$, and also $\rho=\frac{1}{2}\left( \begin{array}{cc} 1/2 & 1/2 \\ 1/2 & 1/2 \\ \end{array}\right)+\frac{1}{2}\left( \begin{array}{cc} 1/2 & -1/2 \\ -1/2 & 12\\ \end{array}\right)$ with $S_{FG}=2$, or other forms. In practice, if the completely mixed state exists without being known how it is prepared, the type-II coarse-graining is necessary by ignoring the entropy contained in the basis which reduces the QFG entropy to von Neumann entropy. This follows from a more general observation that von Neumann entropy is practical in the case that the decomposition of the state matrix $\rho$ is unknown, yet, which can cause some un-physical consequences, such as the negative conditional entropy.

For multi-partite system, the definition of QFG entropy of the whole system is similar with the above, also the QFG entropy for each part follows by type-I coarse-graining, then we can define the mutual information between two parties or among many parties. Particularly, we employ the notation ``(n,m)-mutual information'' for a N-partite state, with $2\leq m\leq n\leq N$, (n,m) means (n-party, m-partition), i.e., the mutual information $I^{(n,m)}$ is defined for n parties and shared in a m-partition way. For instance, $I^{(3,2)}$ for state $\rho_{ABC}$ can be $I(AB:C)$, $I(AC:B)$ or $I(BC:A)$, $I^{(3,3)}$ is just $I(A:B:C)$. Below, let us do some sample calculations. For the Greenberger-Horne-Zeilinger (GHZ) state \cite{ghz}
\begin{equation}\label{}
|\Xi\rangle_{ABC}=\frac{\sqrt{2}}{2}(|000\rangle+|111\rangle),
\end{equation}
the QFG entropies are $S_{FG}(ABC)=1$, $S_{FG}(AB)=$ $S_{FG}(AC)=$ $S_{FG}(BC)=$ $S_{FG}(A)=$ $S_{FG}(B)=$ $S_{FG}(C)=1$, then the (2,2)-mutual information $I(A:B)=$$I(B:C)=$$I(A:C)=1$, the (3,2)-mutual information $I(AB:C)=$$I(BC:A)=$$I(AC:B)=1$, the (3,3)-mutual information $I(A:B:C)=1$. For a general form $|\Xi\rangle=\alpha_1|000\rangle+\alpha_2|111\rangle$, we can find all the entropies are equal to the binary entropy $h(|\alpha_1|^2)$. This unique feature shows that the GHZ state is maximally entangled. The other class of three-qubit entangled state is the W state \cite{w,footnote3}
\begin{equation}\label{}
|W\rangle_{ABC}=\frac{\sqrt{3}}{3}(|001\rangle+|010\rangle+|100\rangle).
\end{equation}
We find that all the entropies are equal to $\log 3$, bigger than the GHZ state. The well-known result for W state is that there is no global entanglement shared by the three parties, yet, here we find that the (3,3)-mutual information is also maximal, the W state is also maximally entangled. A four-partite cluster state \cite{clus} is defined as
\begin{equation}\label{}
|C\rangle_{ABCD}=\frac{1}{2}(|0000\rangle+|0011\rangle+|1100\rangle-|1111\rangle).
\end{equation}
For partition $AB|CD$, $|C\rangle_{AB|CD}=\frac{\sqrt{2}}{2}(|00\rangle_{AB}|\phi^+\rangle_{CD}+$ $|11\rangle_{AB}|\phi^-\rangle_{CD})$, then state $|C\rangle_{AB}=\frac{\sqrt{2}}{2}(|00\rangle+|11\rangle)$ with basis $|00\rangle$ and $|11\rangle$, state $|C\rangle_{CD}=\frac{\sqrt{2}}{2}(|\phi^+\rangle+|\phi^-\rangle)$ with Bell basis $|\phi^{\pm}\rangle$. The QFG entropies are $S_{FG}(|C\rangle_{AB|CD})=1$, $S_{FG}(|C\rangle_{AB})=1$, $S_{FG}(|C\rangle_{CD})=1$, then the mutual information $I_{FG}(|C\rangle_{AB|CD})=1$. The cluster state can also be expressed in partition $AC|BD$ (or same with $AD|BC$), $|C\rangle_{AC|BD}=\frac{1}{2}(|00\rangle_{AC}|00\rangle_{BD}+|01\rangle_{AC}|01\rangle_{BD}+|10\rangle_{AC}|10\rangle_{BD}-|11\rangle_{AC}|11\rangle_{BD})$, $AC$ and $BD$ each forms a four-level system. Then, the QFG entropies are $S_{FG}(|C\rangle_{AC|BD})=2$, $S_{FG}(|C\rangle_{AC})=2$, $S_{FG}(|C\rangle_{BD})=2$, and the mutual information $I_{FG}(|C\rangle_{AC|BD})=2$, different with the above case. We can see that for multi-partite state, the QFG entropies highly depend on the type of partition, which relates to practical situation directly.

For the mutual information (correlation), we have derived some unique results. Firstly, there is no any correlation in single-body state. Secondly, correlation has two types: classical correlation due to mixing, and quantum correlation due to entanglement. Thirdly, pure entangled state only contains quantum correlation. Forth, correlation and the QFG entropy in multi-partite state depend on the type of partition directly. Fifth, the separable state only contains classical correlation, contrary with the result based on quantum discord \cite{discord1,discord2}, as we will see later on.

Furthermore, it is crucial to verify that von Neumann entropy is exactly a kind of quantum coarse-grained entropy. In the original gedanken experiment \cite{neumann}, an ensemble of quantum gas which contains two kinds of particles are stored in a box, immersed in a resevoir keeping constant temperature. The process for the dynamics of the quantum entropy is mainly as follows:

{\em step 1}. Prepare the quantum gas in state $\rho=(1-\lambda)\rho_1+\lambda_2\rho_2$, with $\rho_i=|\psi_i\rangle\langle\psi_i|$ (i=1,2) as the two states of the gas, $\lambda$ specifies the weight (probability) of the two states.

{\em step 2}. Connect with another box with the same volume, use ideal walls to separate the two states into the two boxes, respectively.

{\em step 3}. Get rid of the walls, then compress the two boxes, with the volumes proportional to their weight, to get the original volume of one box. The state is then the same with the initial state.

Based on the above procedure, if the two states $\rho_1$ and $\rho_2$ are distinguishable, the quantum entropy satisfies
\begin{equation}
S(\rho)=(1-\lambda)S(\rho_1)+\lambda S(\rho_2)-\lambda \log (\lambda)-(1-\lambda)\log(1-\lambda),
\end{equation}
note the Boltzmann's constant is ignored. The above relation depends on the decomposition of the state matrix, to avoid this, the entropy of the pure state $|\psi_1\rangle$ and $|\psi_2\rangle$ are set to be the constant zero, then the standard form of von Neumann entropy is derived.

The physical reason that why the entropy should not depend on the decomposition is not clear, though. Actually, in the above argument the technique of coarse-graining is involved. According to type-II coarse-graining, here set $|\psi_1\rangle\equiv|0\rangle$, $|\psi_2\rangle\equiv|1\rangle$, $|0(1)\rangle$ are orthonormal basis, then the state matrix is simply $\rho=\textrm{diag}(1-\lambda, \lambda)$, the resulting entropy is just von Neumann entropy. Furthermore, no matter the $\rm{supp}(\rho_i)$ are orthogonal or not, i.e., whether they can be totally distinguished or not, the QFG entropy relies on the fact that different sectors are prepared unconditionally thus the entropies of sectors can add up together with the mixing entropy.

The technique of coarse-graining and fine-graining, from another view, indicates that the description of entropy, to some extent, depends on our choice. In CM, we also need some choices, e.g., we need to choose the relative zero point for potential. We have to say our choice (observation, measurement) is also physical, which is consistent with the quantum measurement theory. In classical system, we do not need to quantify the quantum entropy although the uncertainty of quantum state exists, i.e., the quantum uncertainty is treated as ``hidden information''. However, for quantum system, the entropy of quantum state cannot be coarse-grained basically, i.e., the QFG entropy depends on the decomposition directly.

\section{BASIC PROPERTIES OF ENTROPY}
\label{sec:3prop}

In this section, we analyze the basic properties of three kinds of entropies: Shannon entropy, von Neumann entropy and the QFG entropy. Note that there exist other kinds of entropy, e.g., Re\'{n}yi entropy, which we do not study here. In the following, we first lay out the primary properties of entropy, then we compare those properties of the three kinds of entropies in details.

We remind that the origin of entropy lies in thermodynamics, where entropy is defined according to the change of heat under a certain temperature. By statistical physics, the macroscopic state is connected to the microscopic state, which leads to the Boltzman entropy. Generally, entropy describes the degree of uncertainty of a random variable, classical or quantum, in sample space. We need to pay attention to the notion of ``random'', which means the underlying dynamics is highly unpredictable. Actually, there does exist the dynamics of the random variable, no matter how complicated it is. The crux is that we ignore the detailed dynamics, which leads to the statistical description where entropy plays the roles as an ``order parameter''.

To define entropy, for simplicity, we take into account the two most primary requirements \cite{wehrl,Nielsen,gc97}, which are (1): the entropy is non-negative; (2): the entropy is additive for independent random variable. Suppose a random variable is $\mathcal{W}\equiv\{W_i, m_i\}$, $W_i$ are the elements, which are not random variable, $m_i$ are the corresponding probabilities, $\sum_i m_i=1$. The entropy $\mathcal{E}$ is defined as
\begin{equation}\label{eq:entropy}
\mathcal{E}(\mathcal{W})=-\sum_i m_i \log m_i.
\end{equation}
It's direct to verify that the definition satisfies the two requirements.

The random variable can be generalized to the case when the elements $W_i$ again are random variable $W_i=\{W^i_{\alpha}, m^i_{\alpha}\}$ with $\sum_{\alpha}m^i_{\alpha}=1$. Then, the definition of entropy is generalized to
\begin{equation}\label{eq:genentropy}
\mathcal{E}(\mathcal{W})=-\sum_i m_i \log m_i+\sum_i m_i \mathcal{E}(W_i),
\end{equation}
and $\mathcal{E}(\mathcal{W}_i)=-\sum_{\alpha} m_{\alpha}^i \log m_{\alpha}^i$. To prove the additivity of the generalized entropy, let us start with the case of two independent random variables. Suppose $\mathcal{W}\equiv\{W_i^a\otimes W_j^b, m_i^a m_j^b\}$, $\otimes$ denotes ``independence'' or ``unconditionality'', the independent random variables $\mathcal{W}^a\equiv\{W_i^a, m_i^a\}$, and $\mathcal{W}^b\equiv\{W_j^b, m_j^b\}$, then the entropy is
\begin{eqnarray}\label{eq:add}\nonumber
\mathcal{E}(\mathcal{W})&=&-\sum_{i,j} m_i^a m_j^b \log (m_i^a m_j^b)+\sum_{i,j} m_i^a m_j^b \mathcal{E}(W_i^a\otimes W_j^b) \\ \nonumber
&=&-\sum_i m_i^a \log m_i^a-\sum_j m_j^b \log m_j^b\\ \nonumber
&& +\sum_i m_i^a \mathcal{E}(W_i^a)+\sum_j m_j^b \mathcal{E}(W_j^b) \\
&=&\mathcal{E}(\mathcal{W}^a)+\mathcal{E}(\mathcal{W}^b),
\end{eqnarray}
which proves the property of additivity. The case of many independent random variables follows directly. Thus, the property of additivity is proved.

From additivity, the property of {\em subadditivity} is induced when the random variables are not independent, i.e., there exists {\em correlation} among them. For the case of two dependent random variables $\mathcal{X}$ and $\mathcal{Y}$, the correlation $\mathcal{I}(\mathcal{X:Y})$, which is called {\em mutual information}, is defined as
\begin{equation}\label{eq:mutual}
\mathcal{I}(\mathcal{X}:\mathcal{Y})\equiv\mathcal{E}(\mathcal{X})+\mathcal{E}(\mathcal{Y})-\mathcal{E}(\mathcal{XY}).
\end{equation}
The primary property of the correlation is that it is non-negative $\mathcal{I}(\mathcal{X}:\mathcal{Y})\geq0$.

The subadditivity property follows the non-negativity of correlation, which is
\begin{equation}
\mathcal{E}(\mathcal{XY})\leq\mathcal{E}(\mathcal{X})+\mathcal{E}(\mathcal{Y}).
\end{equation}
Here, for the quantity $\mathcal{E}(\mathcal{XY})$, the detailed form of the global state is necessary. For instance, when we deal with classical variable, often we need to know the sequence of the signals, like $0,1,1,0,\dots$, not just the probability distribution. For quantum state, given two states $\rho_A$ and $\rho_B$ without the global state, we cannot calculate the mutual information between them. Instead, we can use other quantities to capture the relation between them, such as quantum relative entropy, geometric distance, etc. This manifests that the definition of mutual information depends on both the local and the global states.

Further, from subadditivity, the property of {\em concavity} follows \cite{wehrl}. Suppose two random variables $\mathcal{W}^a=\{W_i, m_i\}$,  $\mathcal{W}^b=\{i, m_i\}$, which together forms $\mathcal{W}=\{W_i\otimes i, m_i\}$. The number $i$ can be viewed as the ``counting'' parameter. Then, the entropy
\begin{eqnarray}\nonumber
\mathcal{E}(\mathcal{W})&=&-\sum_i m_i \log m_i+\sum_i m_i \mathcal{E}(W_i\otimes i) \\ \nonumber
&=&-\sum_i m_i \log m_i+\sum_i m_i \mathcal{E}(W_i) \\
&\leq&-\sum_i m_i \log m_i+\mathcal{E}\left(\sum_i m_i W_i\right),
\end{eqnarray}
where we have used the properties of additivity and subadditivity, which proves the property of concavity
\begin{equation}\label{eq:cancavity}
\sum_i m_i \mathcal{E}(W_i)\leq\mathcal{E}\left(\sum_i m_i W_i\right).
\end{equation}
By comparison of Eqs. (\ref{eq:genentropy}) and (\ref{eq:cancavity}), we can see the difference between $\sum_i m_i \mathcal{E}(W_i)$ and $\mathcal{E}(\sum_i m_i W_i)$ is $-\sum_i m_i \log m_i$, which is sometimes called the ``mixing entropy''. That is, the property of cancavity can also be deduced directly from the definition of entropy since the mixing entropy is non-negative. The formula of entropy in Eq. (\ref{eq:entropy}) describes the process of mixing, and the formula of entropy in Eq. (\ref{eq:genentropy}) describes the process of re-mixing since its elements are themselves random variables.

Relating to mutual information, the {\em conditional entropy} is defined as
\begin{equation}
\mathcal{E}(\mathcal{X|Y})=\mathcal{E}(\mathcal{X})-\mathcal{I}(\mathcal{X}:\mathcal{Y}).
\end{equation}
Obviously, $\mathcal{E}(\mathcal{X|Y})\leq\mathcal{E}(\mathcal{X})$.

There are lots of other properties of entropy especially for multi-random variables case, we do not analyze them in details here.

Next, we turn to the properties of the three entropies, the random variables are: (1) for Shannon entropy: classical random variable $X=\{x_i, p_i\}$; (2) for quantum entropy (including von Neumann entropy and QFG entropy): wave function $|\psi\rangle=\{|i\rangle, \alpha_i\}$, and state matrix $\rho=\{\rho_i, p_i\}$, the properties are studied in details below.

\subsection{Shannon entropy}\label{sec:31shannon}

Shannon entropy plays the central roles in classical information theory, also in quantum information theory. As it is well-known, here we simply present the most common properties without proof.

The Shannon entropy $H(X)$ of single classical random variable $X=\{x_i, p_i\}$, $\sum_i p_i=1$, is defined as
\begin{equation}
H(X)=-\sum_i p_i \log p_i,
\end{equation}
which follows directly from Eq. (\ref{eq:entropy}). It satisfies the properties of non-negativity, additivity, subadditivity, and concavity studied above.

The classical mutual information (correlation) between two random variable $X$ and $Y$ is defined as $I(X:Y)=H(X)+H(Y)-H(X,Y)$, which follows from Eq. (\ref{eq:mutual}). Here, we do not analyze the properties in details.

\subsection{von Neumann entropy}\label{sec:32von}

For quantum entropy, along with the process of mixing, there are another two kinds of operations which can affect entropy: entanglement and measurement. Note that there are some connections and overlaps between the effects of entanglement and measurement. In this section, we analyze the properties of von Neumann entropy and related entropies, e.g., quantum relative entropy.

The von Neumann entropy of the state matrix $\rho$ is defined as
\begin{equation}
S(\rho)=-\textrm{tr}(\rho \log \rho).
\end{equation}
Note that all the properties and the proof of von Neumann entropy can be found in literature, e.g., in Ref. \cite{Nielsen}.\\
{\em Basic properties:}
\newtheorem{Property}{Property}
\begin{Property}
The von Neumann entropy is non-negative, with {\em zero} for pure state.
\end{Property}

The state matrix of pure state can be diagonalized with one diagonal element as ``1'' and others ``0''. The physical reason is that the wave function is said to provide the {\em complete} description of a system.

\begin{Property}
The von Neumann entropy is maximal for completely mixed state.
\end{Property}

By diagonalization, von Neumann entropy reduces to Shannon entropy $H(\lambda_i)=-\sum_i \lambda_i \log \lambda_i$, with $\lambda_i$ as the eigenvalues, $\sum_i \lambda_i=1$. The maximum is reached when all the probabilities $ \lambda_i$ equal, i.e., the state is completely mixed.

\begin{Property}
For $\rho=\sum_i p_i \rho_i$, its von Neumann entropy $S(\rho)=H(p_i)+\sum_i p_i S(\rho_i)$ if $\textrm{supp}(\rho_i)$ are orthogonal.
\end{Property}

The proof can be found in Ref. \cite{Nielsen}. Note that this is the result of the definition of entropy in Eq. (\ref{eq:genentropy}). By comparison with the definition of QFG entropy in Eq. (\ref{qfgrho}), we can see that the condition that the $\textrm{supp}(\rho_i)$ are orthogonal is not necessary for the QFG entropy, since physically the additivity of entropy only requires independence without referring to orthogonality.

The ``joint entropy theorem'' follows from this property, which states that von Neumann entropy for classical-quantum (CQ) state $\rho=\sum_ip_i |i\rangle\langle i|\otimes \sigma_i$ is $S(\rho)=H(p_i)+\sum_ip_iS(\sigma_i)$, where $\{|i\rangle\}$ are orthonormal basis for $A$, $\{\sigma_i\}$ are the sectors for $B$.\\
{\em Property of mixing:}
\begin{Property}
Concavity and mixing:
\begin{equation}
\sum_i p_i S(\rho_i)\leq S\left(\sum_i p_i \rho_i\right)\leq H(p_i)+\sum_i p_i S(\rho_i).
\end{equation}
\end{Property}

The left part follows from the property of concavity in Eq. (\ref{eq:cancavity}), ``='' holds when one of $p_i$ is ``1''. For the right part, the proof can be found in Ref. \cite{Nielsen}; while the right part of the inequality does not always hold \cite{wehrl}.\\
{\em Properties of entanglement:}
\begin{Property}
Subadditivity: For bipartite state $\rho_{AB}$, state of $A$ is $\rho_A$, and state of $B$ is $\rho_B$, then von Neumann entropy satisfies
\begin{equation}
S(\rho_{AB})\leq S(\rho_A)+S(\rho_B).
\end{equation}
\end{Property}

The ``='' holds when $\rho_{AB}=\rho_A \otimes \rho_B$, i.e., the additivity property. The physical reason for subadditivity is due to entanglement or correlation.

\begin{Property}
Strong subadditivity: For tripartite state $\rho_{ABC}$, von Neumann entropy satisfies
\begin{equation}
S(\rho_{ABC})\leq S(\rho_{AB})+S(\rho_{BC})-S(\rho_{B}).
\end{equation}
\end{Property}

The proof for this property is quite complicated \cite{Nielsen,supersub}. On the contrary, we will see below the strong subadditivity for the QFG entropy is quite direct to prove.

The quantum mutual information is defined as $I(\rho_{AB})=S(\rho_A)+S(\rho_B)-S(\rho_{AB})$. With subadditivity, $I(\rho_{AB})\geq0$. The quantum conditional entropy is defined as $S(\rho_A|\rho_B)=S(\rho_A)-I(\rho_{AB})$.

\begin{Property}
Negative conditional entropy: the pure state $|\psi\rangle_{AB}$ for system $A$ and $B$ is entangled iff $S(\rho_A|\rho_B)<0$.
\end{Property}

For pure entangled state $|\psi\rangle_{AB}$, the conditional entropy $S(\rho_A|\rho_B)=-S(\rho_B)$ is negative since $S(\rho_{AB})=0$.

\begin{Property}
Subadditivity of conditional entropy:
\begin{eqnarray}\nonumber
S(\rho_{AB}|\rho_{CD})&\leq&S(\rho_{A}|\rho_{C})+S(\rho_{B}|\rho_{D}), \\ \nonumber
S(\rho_{AB}|\rho_{C})&\leq&S(\rho_{A}|\rho_{C})+S(\rho_{B}|\rho_{C}), \\
S(\rho_{A}|\rho_{CD})&\leq&S(\rho_{A}|\rho_{C})+S(\rho_{A}|\rho_{D}).
\end{eqnarray}
\end{Property}

\begin{Property}
Strong subadditivity of conditional entropy:
\begin{equation}
S(\rho_{ABC}|\rho_{D})+S(\rho_{B}|\rho_{D})\leq S(\rho_{AB}|\rho_{D})+S(\rho_{BC}|\rho_{D}).
\end{equation}
\end{Property}
While this property is not always true for Shannon entropy. The subadditivity and strong subadditivity of conditional entropy follows from the strong subadditivity of von Neumann entropy \cite{Nielsen} from Property $6$.\\
{\em Properties of measurement:}
\begin{Property}
Orthogonal projective measurement increases entropy:
suppose the state $\rho$ under orthogonal projectors $\{P_i\}$ becomes $\rho'$, then
\begin{equation}
S(\rho')\geq S(\rho),
\end{equation}
with equality iff $\rho=\rho'$.
\end{Property}

However, for non-orthogonal projective measurement and positive operator-valued measurement (POVM), the entropy of the system can either increase or decrease depending on the forms of the measurement. The basic observation is that measurement can be viewed as the coupling of the system, which is open, with one apparatus (environment). The change of the entropy of one open system does not exhibit one fixed tendency, which is consistent with the second law of thermodynamics.

We mention that another quantum entropy is the quantum relative entropy defined as $S(\rho||\sigma)=\rm{tr}(\rho \log \frac{\rho}{\sigma})$.
\begin{Property}
Monotonicity of quantum relative entropy:
\begin{equation}
S(\rho_A||\sigma_A)\leq S(\rho_{AB}||\sigma_{AB}).\\
\end{equation}
\end{Property}
Due to the non-negativity and monotonicity of relative entropy above \cite{Nielsen}, it provides an effective way to describe the properties, yet not necessarily entanglement, of quantum state.

\subsection{Quantum fine-grained entropy}\label{sec:33QFG}

By comparison with Shannon entropy and von Neumann entropy, in this section we present the properties of the QFG entropy, some of them are different from those of von Neumann entropy.\\
{\em Basic properties:}
\newtheorem{Property*}{Property*}
\begin{Property*}
The QFG entropy is non-negative, with {\em zero} for the basis of a wave function.
\end{Property*}

As we have discussed, the QFG entropy of pure state is not zero. The pure state lives in the Hilbert space, which can be decomposed into the superposition of a set of orthonormal basis $\{|i\rangle\}$. By comparison with the classical statistical theory, where a random variable $X$ defined as $X=\{x_i, p_i\}$ with different values $x_i$ and probability $p_i$, the basis $|i\rangle$ corresponds to the values $x_i$, which does not contain any entropy. The QFG entropy is defined above in Eq. (\ref{qfg1}), which is one of the main differences from von Neumann entropy.

\begin{Property*}
The QFG entropy of pure state is maximal for balanced superposed state.
\end{Property*}

By ``balanced superposed state'', we mean a pure state $|\psi\rangle=\sum_i \alpha_i |i\rangle$ with the modular of the coefficients $|\alpha_i|$ equal. For the QFG entropy of mixed state, there is no statement for the maximum since it depends on the entropy of the pure states being mixed. Yet, for the classical part of QFG entropy, $H(p_i)$, it takes the maximum when the mixed state is maximally mixed, i.e., all $p_i$ are equal.

\begin{Property*}
Additivity: The QFG entropy of many-body product state $\rho=\rho_1\otimes\rho_2\otimes \cdots$ is the sum of the QFG entropy of each part.
\end{Property*}

We first prove the additivity for bipartite state $\rho=\rho_1\otimes\rho_2$. If $\rho_1=|\psi_1\rangle\langle\psi_1|$, $\rho_2=|\psi_2\rangle\langle\psi_2|$, and $|\psi_1\rangle=\sum_{i=1}^n \alpha_i |e_i\rangle$, $|\psi_2\rangle=\sum_{j=1}^m \beta_j |f_j\rangle$, then the QFG entropy of state $\rho$ is
\begin{eqnarray}\nonumber
S_{FG}(\rho)&=&\sum_{i=1}^n\sum_{j=1}^m|\alpha_i|^2|\beta_j|^2\log(|\alpha_i|^2|\beta_j|^2)\\ \nonumber
&=&\sum_{i=1}^n|\alpha_i|^2\log|\alpha_i|^2+\sum_{j=1}^m|\beta_j|^2\log|\beta_j|^2\\
&=&S_{FG}(\rho_1)+S_{FG}(\rho_2).
\end{eqnarray}

If $\rho_1=\sum_i p_i \rho_i=\sum_i p_i|\psi_i\rangle\langle\psi_i|$, $\rho_2=\sum_j q_j\sigma_j=\sum_j q_j|\phi_j\rangle\langle\phi_j|$, then
\begin{eqnarray}\nonumber
S_{FG}(\rho)&=&S_{FG}(\sum_{i,j}p_iq_j\rho_i\otimes\sigma_j)\\ \nonumber
&=&H(p_iq_j)+\sum_{i,j} p_iq_j S_{FG}(\rho_i\otimes\sigma_j)\\ \nonumber
&=&H(p_i)+H(q_j)+\sum_i p_i S_{FG}(\rho_i)+\sum_j q_j S_{FG}(\sigma_j)\\
&=&S_{FG}(\rho_1)+S_{FG}(\rho_2).
\end{eqnarray}
If sectors $\rho_i$, $\sigma_j$ are further mixed state, it is direct to verify that the additivity still holds. Also, the proof can be extended to the many-body case. Note the method for proof follows from that in Eqs. (\ref{eq:add}).\\
{\em Property of mixing:}
\begin{Property*}
Concavity:
\begin{equation}
\sum_i p_i S_{FG}(\rho_i)\leq S_{FG}\left(\sum_i p_i \rho_i\right).
\end{equation}
\end{Property*}

This inequality follows directly from the definition of the QFG entropy, Eq. (\ref{qfgrho}).\\
{\em Properties of entanglement:}
\begin{Property*}
Subadditivity: For bipartite state $\rho_{AB}$, state of $A$ is $\rho_A$, state of $B$ is $\rho_B$, the QFG entropy satisfies
\begin{equation}
S_{FG}(\rho_{AB})\leq S_{FG}(\rho_A)+S_{FG}(\rho_B).
\end{equation}
\end{Property*}

This property is the same with that of von Neumann entropy, the physical reason is that the mutual information is non-negative.

Relating to Property $7$ of von Neumann entropy, however, the QFG conditional entropy is non-negative. For bipartite pure entangled state $|\psi\rangle_{AB}$, we can see that $S_{FG}(\rho_{AB})=S_{FG}(\rho_A)=S_{FG}(\rho_B)$, then $S(\rho_A|\rho_B)=0$. This result is reasonable, since zero conditional entropy shows that if one party is measured, our uncertainty of the other party is zero, i.e., the two parties are perfectly correlated. We will investigate the problems of entanglement in the next section in details.

\begin{Property*}
Strong subadditivity: For tripartite state $\rho_{ABC}$, the QFG entropy satisfies
\begin{equation}
S_{FG}(\rho_{ABC})\leq S_{FG}(\rho_{AB})+S_{FG}(\rho_{BC})-S_{FG}(\rho_{B}).
\end{equation}
\end{Property*}

From the definition of QFG (3,2)-mutual information $I_{FG}(\rho_{AB}:\rho_C)=$ $S_{FG}(\rho_{AB})+S_{FG}(\rho_C)-$ $S_{FG}(\rho_{ABC})$, then the strong subadditivity reduces to $I_{FG}(\rho_B:\rho_C)\leq I_{FG}(\rho_{AB}:\rho_C)$, which is obvious for the QFG entropy. This is equivalent to the property that conditioning reduces entropy. Note that this proof does not hold for von Neumann entropy, since in which case, the conditional entropy can be negative.

The QFG conditional entropy for bipartite state is defined as $S_{FG}(\rho_A|\rho_B)=S_{FG}(\rho_A)-I_{FG}(\rho_A:\rho_B)$. The properties of subadditivity of QFG conditional entropy are the same as von Neumann entropy, which can be proved by transferring to inequalities of QFG mutual information. However, the strong subadditivity does not always hold for QFG conditional entropy, neither for Shannon entropy, since four-partite QFG mutual information which depends on partition is involved.

{\em Properties of measurement:} Contrary with the property of von Neumann entropy, we find that the orthogonal projective measurement does not necessarily increases entropy. Following the same proof \cite{Nielsen}, the inequality $\rm{tr}(\rho\log\rho)-\rm{tr}(\rho\log\rho')\geq 0$ holds due to the property of relative entropy. The QFG entropy of the initial state $\rho$ is $S_{FG}(\rho)=-\rm{tr}(\rho\log\rho)+S_0$, of the finial state $\rho'$ is $S_{FG}(\rho')=-\rm{tr}(\rho'\log\rho')+S_1$, with constants $S_0$ and $S_1$ depending on the formula of states $\rho$ and $\rho'$, respectively. Then, we can find $S_{FG}(\rho')-S_{FG}(\rho)\geq S_1-S_0$. Since $S_1-S_0$ can be negative, we conclude that the orthogonal projective measurement does not necessarily increase entropy.

Furthermore, the QFG entropy under arbitrary measurement can either increase, decrease, or remains. For instance, during the teleportation process, the information is transmitted between Alice and Bob, and the total amount of information is conserved. Suppose Alice holds the arbitrary state $|\psi\rangle=\alpha_1|0\rangle+\alpha_2|1\rangle$, the QFG entropy is the binary entropy $h(|\alpha_1|^2)$. Bob holds one of the Bell states, the total QFG entropy of the two parties is $2$ bits. After Bell measurement of Alice, she will detect four different results, thus, $2$ bits of information. The states of the rest party will also be of four different forms, yet all have the same entropy $h(|\alpha_1|^2)$. As the result, the total QFG entropy, which is $h(|\alpha_1|^2)+2$ bits, remains unchanged.

We briefly conclude that the three kinds of entropy all satisfy the primary properties of entropy, i.e., non-negativity, additivity, subadditivity, and concavity. The difference is {\em how} they satisfy the properties, and the proof for the properties of QFG entropy follows from rather simple mathematical constructions.

\section{APPLICATIONS OF QUANTUM FINE-GRAINED ENTROPY}
\label{sec:4app}

In this section, we will investigate the possible applications of the method of QFG entropy. For instance, if we assume the whole universe as a pure state, its entropy will be zero, which is apparently conflict with the observation based on thermodynamics. With the method of QFG entropy, the confliction can be resolved. Also, in the schemes of quantum key distribution, namely, the BB84 scheme \cite{bb84}, the bit value $0$ ($1$) is encoded in nonorthogonal polarization states $|H\rangle$ and $|L\rangle$ ($|V\rangle$ and $|R\rangle$) \cite{footnote4}. When the basis defined by $|H\rangle$ and $|V\rangle$ ($|L\rangle$ and $|R\rangle$) is chosen for measurement, the states $|L\rangle$ and $|R\rangle$ ($|H\rangle$ and $|V\rangle$) will show the property of QFG entropy, which is used to detect the eavesdropping. The common argument that measurement will disturb the state is consistent with QFG entropy. One can check that the security of the quantum key distribution is ensured by QFG entropy, and the proof of security remains the same. In the following, we will analyze in details the applications of the method of QFG entropy in entanglement theory, quantum information processing, and quantum thermodynamics.

\subsection{Entanglement theory} \label{sec:41entan}

Entanglement, as well as nonlocality, is viewed as the central character of quantum phenomenon. Many work have been devoted to quantify the degree of entanglement of entangled state. From the view of {\em quantum coherence}, the reason for entanglement also nonlocality is that there exists {\em shared} (or {\em distributed}) {\em coherence} which is supported nonlocally by not only one body of the whole system \cite{footnote5}. Different methods and quantities can be employed based on different physical arguments, e.g., we can quantify the amount of shared coherence (like concurrence \cite{conc}, negativity \cite{negativity}, degree of entanglement \cite{wang2}), the degree of quantum correlation (like entropy of entanglement \cite{bbps,Bennett}, quantum relative entropy \cite{Vedral}, quantum discord \cite{discord1,discord2}), or the degree of nonlocality (like Bell's inequality \cite{bell87}). In this section, we show that, due to the coarse-graining of quantum entropy, some results are inappropriately used in literature. We also need to be careful of the notions of ``entanglement'' and ``quantum correlation'', since they are kind of mixed up in literature in order to quantify the quantum propertiexs of entangled state.

Before our analysis of different measures of entanglement, we define three terminologies which are necessary in case of any confusion. If one quantity can be derived from another quantity, we say these quantities are {\em equivalent}. If several quantities give similar descriptions of the physical reality without confliction, we say these quantities are {\em compatible}. The whole set of compatible quantities are said to be {\em consistent} with the physical reality. For instance, the well-known ``fidelity'' and ``trace distance'' are compatible yet not equivalent, since one cannot always be derived from the other.

For clarity, we also classify two kinds of ``paradigm'' of the methods to quantify entanglement. One is the ``entropy+separablility'' (E-S) paradigm, which quantifies entanglement from the entropic (informational) view, and the entangled state is defined relative to separable state. Within this paradigam, for instance, there are the entanglement of formation and distillation, quantum relative entropy, concurrence, negativity, etc. The other paradigm is based on quantum discord, the ``discord+classicality'' (D-C) paradigm, which quantifies the entanglement from the measurement view, and {\em quantumness} instead of quantum correlation is defined relative to classicality of classical state.

The {\em entropy of entanglement} of the entangled bipartite state $|\Psi\rangle=\sum_i \alpha_i |\psi^i_A\psi^i_B\rangle$ is defined as $E(|\Psi\rangle)=S(\rho_A)=S(\rho_B)=-\rm{tr}(\rho_A\log\rho_A)$.

For example, the entropy of entanglement of Bell states, which are defined as
\begin{eqnarray}
|\psi^{\pm}\rangle &=& \frac{\sqrt{2}}{2}(|01\rangle\pm|10\rangle), \\ \nonumber
|\phi^{\pm}\rangle &=& \frac{\sqrt{2}}{2}(|00\rangle\pm|11\rangle),
\end{eqnarray}
are all the same, e.g., $E(|\psi^+\rangle)=S(\rho_A)=1$, with $\rho_A=\rm{diag}(1/2,1/2)$. The von Neumann entropy of pure state $|\psi^+\rangle$ is $0$, thus, the conditional entropy $S(\rho_A|\rho_B)=-1$, which is not quite intuitive. We know that the Bell state contains one ``ebit'' (bit of entanglement) plus one bit of classical entropy \cite{gpw05}. However, according to the method of QFG entropy, $S_{FG}(|\psi^+\rangle)=S_{FG}(\rho_A)=S_{FG}(\rho_B)=I_{FG}(|\psi^+\rangle)=1$, and $S_{FG}(\rho_A|\rho_B)=0$. The zero conditional entropy means that there is no uncertainty of the unmeasured party conditioned on the measured party, which just shows the two systems are perfectly correlated. The Bell state contains one bit of quantum correlation without any classical correlation. One the contrary, the negative conditional entropy is unreasonable. Although in this case the entropy of entanglement gives the same result with the QFG mutual information, for mixed entangled state, they will provide different results.

The {\em entanglement of formation} (EoF) for bipartite state $\rho_{AB}$ is defined relying on the entropy of entanglement as $E_F(\rho_{AB})=\inf \sum_i p_i E(|\psi_i\rangle)$ for all kinds of decomposition into pure state $\rho_{AB}=\sum_i p_i |\psi_i\rangle\langle\psi_i|$. The EoF quantifies the {\em least} amount of entropy of entanglement to form the entangled state $\rho_{AB}$. According to the QFG entropy, the EoF can be modified to $E_{FG}(\rho_{AB})=\inf \sum_i p_i I_{FG}(|\psi_i\rangle)$.

As an example, we study the Werner state \cite{Werner89} of two parties $A$ and $B$, which is non-separable and can be expressed as
\begin{equation}\label{werner}
\rho_w=(1-z)\frac{{\bm 1}}{4}+z |\psi^+\rangle\langle \psi^+|.
\end{equation}
By coarse-graining, $\rho_A=\rho_B=(1-z)\frac{{\bm 1}}{2}+z \rho_0$, with $\rho_0=\frac{1}{2}\left(\begin{array}{cc}1 & 1 \\ 1 & 1 \\ \end{array}\right)$. The QFG entropy of the three states are
\begin{eqnarray}\nonumber
  S_{FG}(\rho_w) &=& H(z)+(1-z)S_{FG}(\frac{{\bm 1}}{4})+z S_{FG}(|\psi^+\rangle) \\ \nonumber
  &=&H(z)+2-z, \\ \nonumber
  S_{FG}(\rho_A) &=&  H(z)+(1-z)S_{FG}(\frac{{\bm 1}}{2})+z S_{FG}(\rho_0) \\
  &=&H(z)+1=S_{FG}(\rho_B),
\end{eqnarray}
as the result, the mutual information $I_{FG}(\rho_w)=H(z)+(1-z)I_{FG}(\frac{{\bm 1}}{4})+z I_{FG}(|\psi^+\rangle)=S_{FG}(\rho_w)-2 S_{FG}(\rho_A)=H(z)+z$. In the calculation, $S_{FG}(\frac{{\bm 1}}{4})=S(\frac{{\bm 1}}{4})=2$, $S_{FG}(\frac{{\bm 1}}{2})=S(\frac{{\bm 1}}{2})=1$, $I_{FG}(\frac{{\bm 1}}{4})=I(\frac{{\bm 1}}{4})=0$, since we have to do coarse-graining for the noise $\frac{{\bm 1}}{4}$ and $\frac{{\bm 1}}{2}$, and $I_{FG}(|\psi^+\rangle)=1$. In the mutual information $I_{FG}(\rho_w)$, the classical part is $H(z)$, which describes the mixing of Bell state with noise, the quantum part is $z$, which comes from the quantum correlation within the Bell state. Also, we can find $E_{FG}(\rho_w)=z$.

Interestingly, the Werner state can be written in other forms, since we can decompose the noisy term. If it is written as
\begin{eqnarray}\label{werner}\nonumber
\rho'_w&=&\frac{1+3z}{4} |\psi^+\rangle\langle \psi^+|+\frac{1-z}{4} |\psi^-\rangle\langle \psi^-|+ \\
&&\frac{1-z}{4} |\phi^+\rangle\langle \phi^+|+\frac{1-z}{4} |\phi^-\rangle\langle \phi^-|,
\end{eqnarray}
the QFG entropy becomes
\begin{equation}
S_{FG}(\rho'_w)=1-\frac{1+3z}{4}\log\frac{1+3z}{4}-\frac{3-3z}{4}\log\frac{1-z}{4},
\end{equation}
which is different with $S_{FG}(\rho_w)$. And we find that $S_{FG}(\rho'_w)=I_{FG}(\rho'_w)=S_{FG}(\rho'_A)=S_{FG}(\rho'_B)$. This shows that since the Bell states are maximally entangled state, the mixture of Bell states $\rho'_w$ is also maximally entangled, although there exists extra classical mixing entropy.

However, according to the entropy of entanglement, the Werner state, no matter what kind of decomposition, is separable when $z<1/3$. This means that the QFG entropy is not compatible with the entropy of entanglement also the entanglement of formation. On the other hand, the quantum discord of Werner state is non-negative, and increases with $z$ from $0$ to $1$, which seems to be compatible with our result above. The quantum discord is said to capture the quantum correlation properly, yet, we will demonstrate below the notion of quantum discord needs to be reconsidered.

The quantum discord $D(\rho_{\overleftarrow{AB}})$ of a bipartite system is defined as the difference of the total correlation $I(\rho_{AB})$ and the ``classical correlation'' $C(\rho_{\overleftarrow{AB}})$
\begin{eqnarray}
  D(\rho_{\overleftarrow{AB}}) &=& I(\rho_{AB})-C(\rho_{\overleftarrow{AB}}) \\ \nonumber
  &=& S(\rho_B)-S(\rho_{AB})+\max_{\Pi^B_i} \sum_ip_i S(\rho_A^i|\Pi^B_i),
\end{eqnarray}
the arrow means that the measurement $\{\Pi^B_i\}$ is performed on $B$, and the information needs to be transferred from $B$ to $A$. This indicates that the quantum discord is not symmetric, i.e., $D(\rho_{\overleftarrow{AB}})\neq D(\rho_{\overrightarrow{AB}})$, with $D(\rho_{\overrightarrow{AB}})=S(\rho_A)-S(\rho_{AB})+\max_{\Pi^A_i} \sum_i p_i S(\rho_B^i|\Pi^A_i)$. For classical random variable, the classical correlation is equal to the total correlation with zero quantum discord.

The quantum discord relies on the definition of classical correlation \cite{discord1}. For bipartite pure state $\rho_{AB}$, it is shown $D(\rho_{AB})=C(\rho_{AB})=S(\rho_A)=S(\rho_B)=I(\rho_{AB})/2$. This is contrary to our result since it means there exists classical correlation in pure state. The reason is that the quantum discord and classical correlation are defined based on measurement. Due to measurement, the coherence within the state is destroyed, as the result, the quantum entropy transfers to the classical entropy (also see Appendix \ref{app:example}). According to the QFG entropy, $S(\rho_{AB})=S(\rho_A)=S(\rho_B)=I(\rho_{AB})$, and particularly $C(\rho_{AB})=I(\rho_{AB})$, which indicates that the quantity $C(\rho_{AB})$ is quantum, may not be called ``classical correlation''. The next example will verify this point more clearly. It is shown that for the CQ state
\begin{equation}
\rho_{AB}=\sum_i p_i |i\rangle_A\langle i|\otimes \rho_B^i,
\end{equation}
with $\{|i\rangle_A\langle i|\}$ as orthonormal basis of $A$, the classical correlation is equal to the total mutual information, which is $S(\rho_B)-\sum_i p_i S(\rho_B^i)$, since the best measurement on $A$ to gain information about $B$ is the projective measurement with $\{|i\rangle_A\langle i|\}$ \cite{discord1}. However, if the subsystem $B$ is actually composed with another two subsystems, say $CC'=B$, which forms bipartite partition with $A$ as $A|CC'$, and each state $\rho_{CC'}^i$ is a Bell state, labeled as $|\psi^i\rangle\langle\psi^i|$, then the state $\rho_B=\textrm{tr}_A\rho_{AB}$ becomes $\rho_B=\sum_i p_i|\psi^i\rangle\langle\psi^i|$, which can be the Werner state $\rho_w$ studied above. As the result, the classical correlation becomes $S(\rho_w)-\sum_i p_i S(|\psi^i\rangle)=S(\rho_w)$, which can contain entanglement. On the contrary, via QFG entropy, the correlation in the CQ state is totally the classical mixing entropy $H(p_i)$, since CQ state is just a separable state.

Another physical reasoning of quantum discord relies on measurement effect \cite{discord2}. It is stated the classical information is locally accessible, and can be obtained without perturbing the state of the system. Since the separable state can be disturbed by local measurement, the quantum discord of separable state is not zero. For the separable state as defined in Eq. (\ref{eq:separable}) in subsection \ref{sec:23von}, there can be local coherence in each subsystem $A$ and $B$, the way that the local measurement disturbs the whole system is that it will disturb the local coherence, which is different with the distributed coherence, i.e., entanglement. That is, the disturbance of the whole system by the local measurement does not mean there exists quantum correlation. The author in Ref. \cite{Luo} shows that the separable state is the reduced state of a classical state defined in higher dimensional space. It is reasonable that a reduced state of a classical state does not contain any quantum correlation. According to the method of QFG entropy, we have concluded in subsection \ref{sec:23von} that the correlation in separable state is totally classical.

Furthermore, from another point of view, the quantum discord and classical correlation are raised based on the observation that the conditional entropy $S(\rho_A|\rho_B)$ requires the state of $\rho_B$ has been measured. However, this is not the case. The conditional entropy defines the entropy of subsystem $A$ given we have known the {\em existence} of $B$. For instance, for the separable state, our knowledge of the existence of the states of $B$ can be expressed as a set of super-operator $\{\mathcal{L}_i^B\}$, which acts as $\mathcal{L}_i^B(\rho_j^B)=\rho_i^B\delta_{ij}$. The state of $A$ after the action of each $\mathcal{L}_i^B$ is $\rho_i^A$. Then, the conditional entropy is $S(\rho_A|\rho_B)=\sum_i p_i S(\rho_i^A)$, which is equal to $S(\rho_{AB})-S(\rho_B)$, as we have studied in subsection \ref{sec:23von}.

Then, what is the physical meaning of the quantity $C(\rho_{AB})$? Briefly, it is the difference of the entropy $S(\rho_B)$ at the initial time and the average of the entropy of the special state of $B$ after each operation, as $\sum_i p_i S(\rho_B^i)$, at the finial time after the measurement. In Ref. \cite{kw}, it is related to the ``distillable common randomness'' defined as $\lim_{n\rightarrow \infty}C(\rho_{AB}^{\otimes n})$. Further, the classical correlation is related to the entanglement of formation \cite{kw}, and some ``quantum conservation law'' is formed \cite{marcos}. However, these results depends on the coarse-grained entropy, as the result, will not hold anymore in the QFG entropy framework.

The analysis above indicates that the concept of quantum discord is questionable, also is not compatible with the method of QFG entropy. However, the quantum discord has quite wide applications. For instance, the well-known deterministic quantum computing with one pure qubit (DQC1) model \cite{dqc1} shows exponential speedup to solve the normalized trace of a unitary operator. It is verified that the quantum discord plays the central role in this model \cite{Datta}. Based on the method of QFG entropy, it is true there is no quantum correlation between the control qubit and the target ensemble of qubits. However, there exists local coherence within the control qubit. It is not necessary to require the existence of entanglement in all kinds of quantum algorithms, e.g., the BB84 quantum key distribution \cite{bb84}. That is to say, the basic distinction of quantum computing from classical computing is the existence of coherence, either in local or distributed forms. Another remarkable application we want to mention is that the condition for complete-positive, trace-preserving (CPTP) channel is proved to be zero quantum discord of the initial state \cite{cptp}. The role of quantum discord in CPTP channel should be checked, and probably new measure of entanglement or quantumness is needed.

Relating to quantum discord, various formulas are raised, reviewed in Ref. \cite{disreview}, such as thermal discord, measurement-induced disturbance (MID), geometric discord, etc. Strictly speaking, the quantities in the D-C paradigm captures the quantumness, i.e., coherence, instead of entanglement; thus, to link the D-C paradigm with correlation is not suitable. The non-zero quantum discord or, e.g., MID only means there exists coherence (including local and distributed). Instead, we can quantify the quantumness of a state $\rho$ directly, without relating to mutual information. For instance, we can still use some quantities in the D-C paradigm defined without referring to the mutual information to quantify quantumness, such as the geometric discord $D_G(\rho)\equiv \min_\chi ||\rho-\chi||^2$, the relative entropy of discord $D_R(\rho)\equiv \min_\chi S(\rho||\chi)$, where $\chi$ is classical state.

\subsection{Quantum information processing} \label{sec:42QIP}

After the analysis of entanglement above, in this section we study some primary quantities in QIP, which are closely related with entanglement. We find that some quantities, relations, and statements need to be modified due to the method of QFG entropy.

{\em Channel capacity} \cite{holreview}. The classical capacity $C(\mathcal{N})$ of a classical channel $\mathcal{N}$ is defined as
\begin{equation}\label{clacap}
C(\mathcal{N})=\max_{p_x} H(X:Y),
\end{equation}
with the input $X=\{x,p_x\}$, output $Y=\{y,q_y\}$. Note that the input signals are often sent one by one as a sequence in time. From the property of Shannon entropy, $C(\mathcal{N})\leq H(X)$, and $C(\mathcal{N}^{\otimes n})=nC(\mathcal{N})$. When the output is the same with the input, the $C(\mathcal{N})$ reaches the maximal $H(X)$, which means a perfect channel will transmit all the information of the input.

For quantum channel, there exist kinds of capacities. The Holevo classical capacity $\chi(\mathcal{E})$ of a quantum channel $\mathcal{E}$ is defined based on the Holevo quantity $\chi$. Suppose the input is prepared as $\rho=\sum_i p_i \rho_i$, Shannon entropy of the distribution is $H(p_i)$. With the POVM $\{E_j=M_j^{\dagger}M_j\}$, for each operator, the resulted state is $\sigma_j=M_j\rho M_j^{\dagger}/q_j$, $q_j=\textrm{tr}(\rho E_j)$. Shannon entropy of the finial state $\sigma=\sum_j q_j \sigma_j$ is $H(q_j)$. The Holevo bound states that the mutual information between $H(p_i)$ and $H(q_j)$ is bounded as $I(p:q)\leq \chi$, with $\chi=S(\rho)-\sum_i p_i S(\rho_i)$. In the standard approach, if the support of $\rho_i$ is orthogonal, then $\chi=H(p_i)$. Generally $\chi\leq H(p_i)$, then $I(p:q)\leq \chi\leq H(p_i)$. However, according to the method of QFG entropy, it always holds that $\chi=H(p_i)$, then the Holevo bound just describes that $I(p:q)\leq H(p_i)$, which is obvious. The Holevo quantity $\chi$ is not the quantum analogue of classical mutual information. For the Holevo capacity of a quantum channel, which is defined as
\begin{equation}\label{holecap}
 \chi(\mathcal{E})=\max_{\{p_i \rho_i\}}S\left(\mathcal{E}\left(\sum_i p_i \rho_i\right)\right)-\sum_i p_i S\left(\mathcal{E}(\rho_i) \right),
\end{equation}
according to QFG entropy, $\chi(\mathcal{E})=\max_{\{p_i, \rho_i\}} H(p_i)$. We can define a new kind of classical capacity of quantum channel. Suppose the distribution $\{p_i\}$ is encoded into a classical random variable $X=\{i,p_i\}$, and the output distribution $\{q_j\}$ is encoded into another variable $Y=\{j,q_j\}$. Then, $\chi(\mathcal{E})\equiv \max_{\{p_i, \rho_i\}} H(X:Y)$, which is almost the same with $C(\mathcal{N})$. This means the classical capacity of a quantum channel only describes the ``classical'' ability of a quantum channel, without referring to the ability to transmit quantum information.

According to the QFG entropy, even the input and output states of a quantum channel are the same, the QFG entropy can be different when their decompositions are different. Also, even we know the decompositions of the input and output states, we cannot calculate the quantum mutual information between them if we do not know the global state, which is yet the case in practice. We can define the quantum capacity $Q(\mathcal{E})$ relying on geometric distance, e.g., fidelity. For instance, the quantum capacity $Q(\mathcal{E})$ of a quantum channel $\mathcal{E}$ can be defined as
\begin{equation}\label{quancap}
Q(\mathcal{E})=\max_{\{p_i, \rho_i\}} S_{FG}(\rho) \left(\rm{tr}\sqrt{\sqrt{\rho}\mathcal{E}(\rho)\sqrt{\rho}}\right)^2.
\end{equation}
When the output $\mathcal{E}(\rho)$ equals to the input $\rho$, the fidelity is one, then the quantum capacity reaches its maximal $S_{FG}(\rho)$, which means that the perfect quantum channel will transmit all the information, including classical and quantum, of the input. It is easy to see that $Q(\mathcal{E})$ is additive, no regularization is needed.

There are other quantum capacities in the literature \cite{coher,lloyd,private,asistant}. The evolution of the input $\rho$ under the channel with output $\sigma$ can be viewed as one unitary operation on the input and one ancilla $E$, the {\em coherent information} \cite{coher,lloyd} is defined as $\mathcal{I}(\mathcal{E})=\max_{\{p_i, \rho_i\}} (S(\sigma)-S(\rho_E))$, whose regularization gives the quantum capacity $\mathcal{Q}(\mathcal{E})$. Recently, study showed that two zero quantum capacity channels can make a positive quantum capacity channel, called ``superactivation'' \cite{smith,bran}. Here we do not intend to analyze this problem in detail. We would like to point out that the notion of coherent information is not compatible with the quantum capacity we defined above. Although the coherent information is smaller than the entropy of input state, it can be negative, as originally noted \cite{coher}. On the contrary, quantum capacity $Q(\mathcal{E})$ we defined is never negative and bounded by the input entropy, thus can be a measure of quantum capacity.

{\em Additivity of entanglement of formation etc.} It is proved by Shor that four conjectures are equivalent: the additivity of minimal entropy output of a quantum channel, the additivity of the Holevo capacity of a quantum channel, the additivity of the entanglement of formation, and the strong superadditivity of the entanglement of formation, see Ref. \cite{shor} and references therein. Here we study this problem from the method of QFG entropy. First, for the entanglement of formation, we have pointed out that in the last subsection it needs to be modified to $E_{FG}(\rho_{AB})=\inf I_{FG}(AB)$, with $I_{FG}(AB)=\sum_i p_i I_{FG}(|\psi_i\rangle)$. We can easily prove that $E_{FG}$ is additive. Note the $I_{FG}$ for four-partite state needs to be the (4,2) partition. Suppose one four-partite state $\varpi_{ABCD}=\rho_{AB}\otimes\sigma_{CD}$, then following the definition of QFG mutual information, we have $I_{FG}(AC:BD)=$$I_{FG}(AD:BC)=$$I_{FG}(AB)+I_{FG}(CD)\equiv I_{FG}(ABCD)$. Now suppose there exist decompositions $\rho_{AB}=\sum_i p_i \rho_i$ and $\sigma_{CD}=\sum_j q_j \sigma_j$ minimize $I_{FG}(AB)$ and $I_{FG}(CD)$, respectively, then obviously $I_{FG}(ABCD)$ is also minimized. Conversely, suppose $I_{FG}(ABCD)$ is minimized by the set $\{p'_i, \rho'_i\}$ and $\{q'_j, \sigma'_j\}$, then the two set must also minimize $I_{FG}(AB)$ and $I_{FG}(CD)$ respectively, due to the linearity of the function. As the result, $I_{FG}(ABCD)$ and $I_{FG}(AB)+I_{FG}(CD)$ are minimized by the same decompositions, i.e., the modified entanglement of formation is additive. The strong superadditivity of the entanglement of formation follows directly due to the definition of QFG mutual information. The Holevo capacity of a quantum channel is also modified in the study above, which is obviously additive. It can be checked that the proof of the equivalence of the four conjectures are compatible with the method of QFG entropy. Recently, it was shown the minimal entropy output is not additive \cite{hastings}, yet no concrete example has been constructed, also it is shown later the minimal entropy output is locally additive \cite{gour}. Their proof should hold on their own right since the method of QFG entropy is not involved. Our argument above could be improved by more detailed study. In addition, relating to entanglement distillation protocols, the entanglement cost is defined as the regularization of entanglement of formation, namely $E_c=\lim_{n\rightarrow \infty}\frac{E_F(\rho_{AB}^{\otimes n})}{n}$. Since the property of additivity after modification, the entanglement cost is equal to the entanglement of formation.

\subsection{Quantum thermodynamics} \label{sec:43statis}

Entropy was originally termed in classical thermodynamics. It seems the definition of thermodynamic entropy relies on temperature, which is a macroscopic (classical) parameter. Since entropy is connected with statistics also quantum statistics, how about temperature? Recently, quantum thermodynamics research, e.g., quantum heat engines \cite{heateng1,heateng2,heateng3,quan}, Maxwell's demon \cite{logical1,logical2,logical3,thermo-cord,demonrev}, have achieved lots of progress. A quantum Carnot engine was proposed via a single particle trapped in 1D infinite square potential well, the efficiency was shown the same as the Carnot efficiency \cite{heateng1}. Interestingly, it was noted the entropy of the particle wave function needs to be the QFG entropy, yet without any physical argument. To make the similarity of quantum and classical engines more clearly, the analog of temperature was further introduced for the single pure quantum state \cite{heateng2}, and the formula for efficiency is exactly the same with the Carnot efficiency, namely, $\eta=1-\frac{T_c}{T_h}$. To our best knowledge, these work can be viewed as the first applications of the QFG entropy in literature. Various quantum heat engines are investigated in details in Ref. \cite{quan}, the quantum isothermal, isoentropic, isochoric, and isobaric processes are discussed, here we do not focus on this study.

Next, we turn to present our approach to define the concept of quantum temperature. For the general quantum state matrix $\rho=\sum_i p_i |\psi_i\rangle\langle\psi_i|$, $|\psi_i\rangle=\sum_k \alpha_{ki} |k_i\rangle$, the QFG entropy is $S_{FG}(\rho)=H(p_i)+\sum_i p_i S_{FG}^i$, with $S_{FG}^i=-\sum_k \lambda_{ki}\log \lambda_{ki}$, and $\lambda_{ki}\equiv |\alpha_{ki}|^2$. Assume the Hamiltonian is $\mathcal{H}$, then the energy of the state $\rho$ is $U=\textrm{tr}(\rho\mathcal{H})=\sum_i p_i E_i$, with $E_i\equiv\langle\psi_i|\mathcal{H}|\psi_i\rangle=\sum_k e_{ki} \lambda_{ki}$.

The variation of the energy is
\begin{eqnarray}\label{eq:firstlaw}\nonumber
  \delta U&=& \sum_i  E_i\delta p_i +\sum_i  p_i\delta E_i \\ \nonumber
   &=&  \sum_i  E_i\delta p_i +  \sum_i p_i \sum_k  e_{ki}\delta \lambda_{ki}+\sum_i p_i \sum_k  \lambda_{ki}\delta e_{ki} \\
   &\equiv& \delta Q+ \delta W,
\end{eqnarray}
with the heat $ \delta Q\equiv\sum_i  E_i\delta p_i +  \sum_i p_i \sum_k  e_{ki}\delta \lambda_{ki}$, the work $\delta W\equiv\sum_i p_i \sum_k  \lambda_{ki}\delta e_{ki}$. Note two terms contribute to the change of heat, the first term is the common classical one, denoted as $\delta Q_c$, the second term is the quantum one, set $\delta Q_{FG}^i\equiv\sum_k  e_{ki}\delta \lambda_{ki}$. The equations (\ref{eq:firstlaw}) just express the quantum first law of thermodynamics.

The variation of the QFG entropy is
\begin{equation}
    \delta S_{FG}(\rho)=H(\delta p_i)+\sum_i S_{FG}^i\delta p_i +\sum_i p_i \delta S_{FG}^i,
\end{equation}
denote the classical part as $\delta S_c\equiv H(\delta p_i)$. The definition of quantum temperature follows directly. The classical temperature, which is coarse-grained, is $T_c\equiv\frac{\delta Q_c}{\delta S_c}$, the QFG temperature is $T_{FG}^i\equiv\frac{\delta Q_{FG}^i}{\delta S_{FG}^i}$, and the quantum temperature can be derived as
\begin{eqnarray}\label{eq:qutem}\nonumber
T_q&\equiv&\frac{\delta Q}{\delta S_{FG}}\\ \nonumber
&=&\frac{\sum_i  E_i\delta p_i +  \sum_i p_i \sum_k  e_{ki}\delta \lambda_{ki}}{H(\delta p_i)+\sum_i S_{FG}^i\delta p_i +\sum_i p_i \delta S_{FG}^i}  \\
&=&\frac{H(\delta p_i)T_c+ \sum_i T_{FG}^ip_i\delta S_{FG}^i}{H(\delta p_i)+\sum_i S_{FG}^i \delta p_i+ \sum_i p_i\delta S_{FG}^i}.
\end{eqnarray}
The quantum temperature $T_q$ can be reduced to the other two under certain limits. If the state $\rho$ is pure, it is direct to check that $T_q=T_{FG}^i=T_{FG}$, there is no classical contribution for the temperature. If we employ coarse-graining of the state, i.e., set $S_{FG}^i$, $\delta S_{FG}^i$ as zero, then $T_q=T_c$, there is no quantum contribution for the temperature. The study of quantum heat engine \cite{heateng2} demonstrated that $T_c$ and $T_{FG}$ play the similar roles since they have similar formula. Our analysis of quantum temperature indicates that temperature as well as entropy relies on the feature of statistics, no matter classical or quantum.

In addition, one point to note is that the coherence of the pure single quantum state does not enter into the formula of temperature directly, which indicates that a mixed state can also perform as the work substance of the quantum heat engine. Coherence relates to the way of transition between the eigenstates $|k_i\rangle$ (via photon, phonon etc), for which the classical counterpart is the scattering among molecules in the gas; and coherence also relates to the geometric phase effect during the cycle, which is interesting for further study.

Another focus of research centers around the second law of thermodynamics, for instance, to relate entanglement with the second law \cite{deficit,Popescu,Brandao,Kim,negentropy}. For open system the entropy can decrease, which is not stated directly by the second law. As we have shown in section \ref{sec:33QFG}, the QFG entropy can either increase or decrease for the quantum open system. The entanglement cannot increase under LOCC, namely, local operation and classical communication, which shows {\em irreversibility}. Treating entanglement as distributed coherence, this irreversibility means that this kind of coherence cannot emerge spontaneously. The underlying issue is the ``conservation of coherence'', as well as entropy and energy, which is not so clear at present. Also, some quantities are introduced to describe the quantum effects relating with thermodynamics, e.g., thermal discord \cite{thermo-cord}, work deficit \cite{deficit}, however, based on rough observation, these quantities rely on von Neumann entropy, as the result, are not compatible with the method of QFG entropy.

\section{CONCLUSION}
\label{sec:5CONCLUSION}

In this work, we explored the method of quantum fine-grained (QFG) entropy, including its physical foundation, properties, and potential applications. Our results based on the QFG entropy are quite different with those related work in literature. For clarity, we make one brief summary.

We classified two types of coarse-graining, type-I and type-II, von Neumann entropy employs the latter type. The QFG entropy of pure state is not zero, since the complete description of wave function is inherent statistical. Different decompositions of the same $\rho$ contain different quantum information, based on our study of the formalism of state matrix. The QFG entropy satisfies the primary properties of entropy, namely, non-negativity, additivity, subadditivity, concavity, also strong subadditivity, we note that more systematic analysis might be needed.

For the applications, we investigated the concept of entanglement, which is actually a kind of coherence instead of information, from the view of QFG entropy. Some problems can be resolved, e.g., we showed that the conditional entropy cannot be negative, the separable state contains only classical correlation, the concept of quantum discord is questionable. Furthermore, we verified that the conjecture for the additivity of the entanglement of formation, the minimal entropy output, the Holevo capacity of a quantum channel, and the strong superadditivity of the entanglement of formation is valid, after some modifications due to the formula of QFG entropy. Relating to thermodynamics, we found that the laws of thermodynamics can be generalized to the quantum regime, particularly, the quantum temperature can be defined based on the QFG entropy, the irreversibility of entropy and entanglement can be explored also with the method of QFG entropy.

The quantum fine-grained entropy has quite natural physical foundations, and makes the characters of quantum information much easier to understand. We expect that there can be more investigations and applications in the future.

\appendix

\section{Entanglement without local mixing}
\label{app:example}

Here we show that in entangled state, the local system does not necessarily stay as mixed state. Consider the theoretical situation: two qubit atoms $A$ and $B$ are coupled by one photon, which can be emitted by one atom then absorbed by the other. There is no decoherence with the environment or vacuum. Then, the population of the ground (excited) state of $A$ behaves the same with the population of the excited (ground) state of $B$, i.e., Rabi oscillation. Thus, the two atoms are in resonance with each other, i.e., in entangled state. Obviously, each of the atom is in superposed state. For simplicity, suppose the entangled state is $|\psi\rangle=\frac{\sqrt{2}}{2}(|e_Ag_B\rangle-|g_Ae_B\rangle)$. Then, state $A$ is $|\psi_A\rangle=\frac{\sqrt{2}}{2}(|e_A\rangle-|g_A\rangle)$ by ignoring $B$, state $B$ is $|\psi_B\rangle=\frac{\sqrt{2}}{2}(|g_B\rangle-|e_B\rangle)$ by ignoring $A$.

The QFG entropies can be calculated as $S_{FG}(|\psi\rangle)=S_{FG}(|\psi_A\rangle)=S_{FG}(|\psi_B\rangle)=I_{FG}(|\psi\rangle)=1$. According to the common approach, the state of $A$ ($B$) is derived by tracing out system $B$ ($A$), as $\rho_A=\rho_B=\rm{diag}(1/2,1/2)$. Then, QFG entropies can be calculated as $S_{FG}(|\psi\rangle)=S_{FG}(\rho_A)=S_{FG}(\rho_B)=I_{FG}(|\psi\rangle)=1$, which is the same with our approach. We see that tracing turns the quantum entropy $S_{FG}(|\psi_{A,B}\rangle)$ into the classical entropy $S_{FG}(\rho_{A,B})$, yet, the amount of entropy remains. Indeed, tracing is a kind of quantum measurement with the optimal projectors, i.e., it decohers the coherence of the superposed state of the sub-systems leading to the mixture. If we measure the states of atoms $A$ or $B$, the local coherence will be destroyed. Generally, we will always observe mixed state if we measure the sub-system, however, this does not mean before the measurement, the sub-system stays as mixed state. Relating to the problem studied here, Ref. \cite{wang1} provided a physical way for the classification of quantum state.

\section{Superposition paradox}
\label{app:para}

Here we raise a problem, which might be noticed widely yet not formed as a standard paradox. The problem is that the superposition of superposed state can lead to non-superposed eigenstate! For instance, $|\phi_1\rangle=(|0\rangle+|1\rangle)/\sqrt{2}$, $|\phi_2\rangle=(|0\rangle-|1\rangle)/\sqrt{2}$, then $(|\phi_1\rangle+|\phi_2\rangle)/\sqrt{2}=|0\rangle$. The question is that: is there coherence within state $|0\rangle$? Note that here $|0(1)\rangle$ can also stands for many-body state, e.g., $|0(1)\rangle\equiv|00(11)\rangle$, then we can also ask: is there entanglement within state $|0\rangle$?

The paradox comes from the relation between reality and decomposition of state matrix. If states $|\phi_1\rangle$ and $|\phi_2\rangle$ are superposed together, i.e., they exist in reality, they are treated as eigenstates: $|\phi_1\rangle\equiv|+\rangle$, $|\phi_2\rangle\equiv|-\rangle$, which cannot be written as superposition of $|0\rangle$ and $|1\rangle$ again, since in this situation states $|0\rangle$ and $|1\rangle$ only potentially exist due to the property of propensity of quantum state \cite{wang1}. Then, $|0\rangle=(|+\rangle+|-\rangle)/\sqrt{2}$ is superposed state under basis $|\pm\rangle$. On the contrary, we also say $|\pm\rangle$ is superposed state under basis $|0(1)\rangle$. Superposition and entanglement relate to the representation we choose. The physical significance of basis is consistent with the method of quantum fine-grained entropy, the superposition paradox can thus be resolved.


\begin{thebibliography}{99}
\bibitem{footnote1} Of course, the quantum dynamics can also be described in other kinds of space, e.g., the quantum phase space based on the Wigner distribution function. The corresponding quantum coarse-graining should be apparently different.
\bibitem{footnote2} Note that there also exists the origin for the mixing of $\mathcal{H}_I$ (or $\mathcal{H}_I'$). In this case, the coherence frame is field, with which the many-body system form $\rho=\sum_i p_i \rho_i\otimes \sigma_i$, tracing out the field is characterized as ``subdecoherence'' \cite{wang1}. This physical picture does not affect the conclusion of this paper.
\bibitem{footnote3} By type-I coarse-graining of parties $AB$, the state of party $C$ will be $|C\rangle=\frac{\sqrt{3}}{3}(|1\rangle+|0\rangle+|0\rangle)$, note that we cannot sum up the two $|0\rangle$ since they belong to different brunches. A more comfortable way to calculate entropy is to use trace, as tracing does not change the amount of QFG entropy except it converts the quantum entropy into classical entropy.
\bibitem{footnote4} The symbols $H$, $V$, $L$, $R$ have their usual meanings, i.e., horizontal, vertical, left-hand circular, right-hand circular, respectively.
\bibitem{footnote5}By a rough comparison, the electron in metal can be bounded or itinerant, use the language of coherence, viewing the electron as a ``coherence package'', we can say the electron can be in local state or entangled (nonlocal) state.
\end{thebibliography}
\end{document}